\documentclass[twocolumn,showpacs,preprintnumbers,superscriptaddress]{revtex4}
\usepackage{mathrsfs}
\usepackage{amssymb}
\usepackage{amsmath}
\usepackage{graphicx}
\usepackage{dcolumn}
\usepackage{bm}
\usepackage{graphicx}
\usepackage{float}
\usepackage[normalem]{ulem}
\usepackage{color}

\begin{document}
\title{Searching for New Spin-Velocity Dependent Interactions by Spin Relaxation of Polarized $^{3}$He Gas   }
\author{H. Yan}
\email[Corresponding author: ]{hyan@caep.cn}\affiliation{Key Laboratory of Neutron Physics, Institute of Nuclear Physics and Chemistry, CAEP, Mianyang, Sichuan, 621900,China }

\author{G.A. Sun}
\affiliation{Key Laboratory of Neutron Physics, Institute of Nuclear Physics and Chemistry, CAEP, Mianyang, Sichuan, 621900,China }

\author{S.M. Peng}
\affiliation{Institute of Nuclear Physics and Chemistry,CAEP, Mianyang, Sichuan, 621900,China }

\author{Y. Zhang}
\affiliation{Key Laboratory of Neutron Physics, Institute of Nuclear Physics and Chemistry, CAEP, Mianyang, Sichuan, 621900,China }
\affiliation{School of Nuclear Science and Technology, University of Science and Technology of China, Hefei,  230026, China}

\author{C. Fu}
\affiliation{Department of Physics, Shanghai Jiaotong University, Shanghai, 200240, China }

\author{H. Guo}
\affiliation{Department of Physics, Southeast University, Nanjing, 211189, China }

\author{B.Q. Liu}
\affiliation{Key Laboratory of Neutron Physics, Institute of Nuclear Physics and Chemistry, CAEP, Mianyang, Sichuan, 621900,China }

\date{\today}
\begin{abstract}
We have constrained possible new interactions which produce nonrelativistic potentials between polarized neutrons and unpolarized matter proportional to  $\alpha\vec{\sigma}\cdot\vec{v}$ where $\vec{\sigma}$ is the neutron spin and $\vec{v}$ is the relative velocity. We use existing data from laboratory measurements on the very long $T_{1}$ and $T_{2}$ spin relaxation times of polarized $^{3}$He gas in glass cells.
Using the best available measured $T_{2}$ of polarized $^{3}$He gas atoms as the polarized source and the earth as an unpolarized source, we obtain constraints on two new interactions. We present a new experimental upper bound on possible vector-axial-vector($V_{VA}$) type interactions for ranges between $1\sim10^{8}$m. In combination with previous results, we set the most stringent experiment limits on $g_{V}g_{A}$ ranging from $\sim\mu$m to $\sim10^{8}$m. We also report what is to our knowledge the first experimental upper limit on the possible torsion fields induced by the earth on its surface. Dedicated experiments could further improve these bounds by a factor of $\sim100$. Our method of analysis also makes it possible to probe many velocity dependent interactions which depend on the spins of both neutrons and other particles which have never been searched for before experimentally.  
\end{abstract}
\maketitle
In recent years, various models of new physics beyond the Standard Model have been studied in which new massive particles such as the axion, familon and majoron, {\it etc.} were theoretically introduced \cite{PDG14}.
New macroscopic interactions meditated by WISPs (weakly-interacting sub-eV particles) have also been theoretically proposed. The interaction ranges of these new forces range from nanometers to astronomical distance scales. The fact that the dark energy density is on the order of (1 meV)${^4}$  corresponding to a length scale of $~100$ $\mu$m  also encourages people to
 search for new physical phenomena around this scale \cite{ADE09}. Various experiments have been performed or proposed recently to search for a subset of these new interactions which could couple to the spin of the neutron/electron. Polarized neutron beams were used in the experiments presented in  Refs.\cite{PIE12,YAN13}. Polarized noble gases were used in Refs.\cite{ZHE12,BUL13,TUL13,CHU13}. Atomic magnetometers were applied in Refs.\cite{VAS09,HUN13}. Experimental schemes using polarized atom beams have been recently proposed in Ref.\cite{YAN14}.

The idea that exotic new interactions might be spin/velocity dependent is quite fascinating. For example,  the theoretically proposed photinos interact with nuclei only through spin-dependent forces~\cite{WIT85}. Ref.\cite{DOB06} analyzed the possible nonrelativistic potentials between spin-$1/2$ fermions from spin $0$ and spin $1$ boson exchange and found 16 possible new interactions, 10 of  which depend both on the spin states and the relative velocity between particles. Torsion, a twisting of spacetime coupled to intrinsic spin, which has been included in many models which extend general relativity~\cite{HEH76,SHA02}, can also induce  spin-velocity dependent interactions between an unpolarized source and the spin. Hari-Dass interaction~\cite{HAR76} is another example of spin-velocity dependent interactions. These possible new interactions have eluded detection so far. It might be that the new interactions are either too weak to be detected, or the spin/velocity dependence causes difficulties for the conduct of sensitive experiments. 

Either for the torsion-induced or the vector-axial-vector ($V_{VA}$) type interaction, the spin of polarized noble gases like $^{3}$He can interact with an unpolarized source through a potential proportional to
\begin{equation}\label{eqn.sdv}
V=\alpha\vec{\sigma}\cdot\vec{v},
\end{equation}
where $\vec{\sigma}$ are the Pauli matrices and $\vec{v}$ is the relative velocity between the spin and the unpolarized source.
The parameter $\alpha$ has dimensions of momentum and depends on factors specific to the interaction such as the probe to source distance, the source mass or the nucleon number, {\it etc.} Since it is known that the spin of the $^{3}$He nucleus is dominated by the spin of its unpaired neutron~\cite{FRI90}, in this paper we interpret constraints on $\alpha$ in terms of neutron properties. To help develop intuition for the physical effects of $V$ it is useful to imagine that $V$ produces an effective pseudo-magnetic field:
\begin{equation}
V=\vec{\mu}\cdot\vec{B'},
\end{equation}
where $\vec{\mu}$ is the magnetic moment,  $\gamma$ the gyromagnetic ratio of the spin polarized particle, and $\vec{B'}={2\alpha\vec{v}}/{\hbar \gamma}$ the pseudo-magnetic field. This pseudo-magnetic field is along the $\vec{v}$ direction and has a strength of ${2\alpha v}/{\hbar \gamma}$. Searching for the new physics from this potential is then equivalent to detecting the pseudo-magnetic field $\vec{B}^{'}$.

Spin polarized neutron/atom beams are convenient to probe these spin-velocity dependent interactions since a large
relative velocity between the probe and the source can be easily realized. However, the number of the probe particles is limited by the phase space density of the beam. Larger phase space densities of polarized probe
particles can be obtained by using ensembles of polarized gases, but the polarized noble gas ensembles which can support sensitive NMR measurements of the spin dynamics needed for this search are usually sealed in glass cells. It would be technically difficult to realize a large relative velocity between the source mass and the probe particles inside delicate glass cells.

Though $\langle \vec{v}\rangle$ is zero for atoms of the glass sealed noble gas, $\langle {v}^{2}\rangle$ is not. The nonzero $\langle {v}^{2}\rangle$ in the presence of a $\vec{\sigma}\cdot\vec{v}$ type interaction will change the spin relaxation times of polarized noble gases. Although it is a second order effect, in this case there is no need for bulk motion of either the polarized or unpolarized masses in the experiments. Thus it is possible to detect or constrain the new physics by the longitudinal spin relaxation time ($T_{1}$) or the transverse relaxation time ($T_{2}$) of polarized noble gases. Here $T_{1}$ refers to the mean time for a spin polarized  ensemble to return to its thermal equilibrium state and $T_{2}$ the mean time that polarized spins to lose coherence when processing along the longitudinal main field while the polarization is tipped to the transverse direction~\cite{SLI89}.   For the best available $T_{1}$ \cite{POK10} and $T_{2}$ \cite{PET10,GEM10} data, our research indicates that the constraint on $\alpha$ from $T_{2}$ is tighter than that from $T_{1}$. In what follows, we will first describe how the $\alpha\vec{\sigma}\cdot\vec{v}$ interaction affects the spin relaxation times of the polarized $^{3}$He gas, then we will constrain $\alpha$ by using the best available $T_{2}$ measured in the experiment. Furthermore, by using this constraint of $\alpha$ and the earth as a source, we obtain new limits on two different types of new interactions,  vector-axial-vector interaction ($V_{VA}$) and a linear combination of the time component of possible torsion fields from the earth.

\section{ Constraining $\alpha$ by $T_{1}$ and $T_{2}$ of spin polarized $^{3}$He gas}
Highly polarized, dense ensembles of polarized $^{3}$He gas have been developed over the last few decades for scientific applications in nuclear/particle physics, neutron spin filters and medical imaging~\cite{HAP72,RIC02,WAL97}.
There are already some examples using the spin relaxation time to constrain the scalar-pseudo-scalar type ~\cite{MOO84} interaction which is spin dependent. $T_{1}$ is used in Ref.\cite{POK10} while $T_{2}$  in Refs.\cite{PET10,FU11}. Assume the magnetic field is along $\hat{z}$, then using the Redfield theory~\cite{MCG89,PET10}, the longitudinal and transverse relaxation times of the polarized $^{3}$He gas due to a randomly fluctuating magnetic field can be expressed as:
\begin{eqnarray}
& &\Gamma_{1}=\frac{1}{T_{1}}=\frac{\gamma^{2}}{2}[S_{Bx}(\omega_{0})+S_{By}(\omega_{0})],\nonumber\\
& &\Gamma_{2}=\frac{1}{T_{2}}=\frac{\gamma^{2}}{4}[S_{Bx}(\omega_{0})+S_{By}(\omega_{0})+2S_{Bz}(0)],
\end{eqnarray}
where
\begin{equation}
S_{Bx}(\omega_{0})=\int^{+\infty}_{-\infty}\langle B_{x}(t)B_{x}(t+\tau)\rangle e^{-i\omega_{0}\tau} d\tau.
\end{equation}
Here $\langle ...\rangle$ represents the ensemble average, $B_{x}$ is the $\hat{x}$ component of the fluctuating magnetic field as seen in the rest frame of the diffusing polarized $^{3}$He atom, and $\omega_{0}=\gamma B_{0}$ is the Larmor frequency. Using above formulas, it is easy to see that the pseudo-magnetic field induced by the $\alpha\vec{\sigma}\cdot\vec{v}$  interaction will change $T_{1}$ as follows:
\begin{eqnarray}\label{eqnT1}
\Gamma_{1}=\frac{1}{T_{1}}=\frac{4\alpha^{2}}{\hbar^{2} }\int^{+\infty}_{-\infty}\langle v_{x}(t)v_{x}(t+\tau)\rangle e^{-i\omega_{0}\tau} d\tau.
\end{eqnarray}
When the Lamor frequency $\omega_{0}$ is much larger than $1/\tau_{D}$ with $\tau_{D}\propto{L^{2}}/{D}$ ($L$ is the characteristic length of the gas sealed cell in the dimensions transverse to the magnetic field), the autocorrelation function for the  velocity is \cite{MCG89}:
\begin{equation}\label{eqnVcor}
\langle v_{x}(t)v_{x}(t+\tau)\rangle=\langle v_{x}^{2}\rangle e^{-\frac{|\tau|}{\tau_{c}}}=\frac{1}{3}\langle v^{2}\rangle e^{-\frac{|\tau|}{\tau_{c}}},
\end{equation}
where $\tau_{c}$ is the average collision time of the atoms.

Plugging in the data given in Ref.\cite{POK10}, $\tau_{c}=3\times10^{-10}$s, $T_{1}^{\textrm{rem}}=2664$\textrm{h} ($1\sigma$ value), $\omega_{0}=10^{5}\textrm{s}^{-1}$, we obtain an upper limit to $\alpha$ as:
\begin{equation}\label{eqn.aT1}
\alpha\leq7.7\times10^{-37}\textrm{kg}\cdot \textrm{m}\cdot \textrm{s}^{-1}.
\end{equation}
Similarly, the relaxation time $T_{2}$ caused by the $\alpha\vec{\sigma}\cdot\vec{v}$ type interaction  can be expressed as:
\begin{eqnarray}\label{eqnT2}
\frac{1}{T_{2}}&=&\frac{\alpha^{2}}{\hbar^{2} }\int^{+\infty}_{-\infty}[\langle v_{x}(t)v_{x}(t+\tau)+v_{y}(t)v_{y}(t+\tau)\rangle e^{-i\omega_{0}\tau}\nonumber\\
 & &\qquad\quad+2\langle v_{z}(t)v_{z}(t+\tau)\rangle]d\tau.
\end{eqnarray}
Here we need to calculate $\langle v_{x}(t)v_{x}(t+\tau)\rangle$ under the condition that the magnetic field is small such that Eqn.(\ref{eqnVcor}) cannot be applied anymore. According to Refs.\cite{SQU77,GOU10}, we can derive
\begin{eqnarray}
\int^{+\infty}_{-\infty}\langle v_{x}(t)v_{x}(t+\tau)\rangle e^{-i\omega_{0}\tau} d\tau=\omega_{0}^{2}S_{x}(\omega_{0}),
\end{eqnarray}
where $S_{x}(\omega_{0})$ is defined as the Fourier transformation of $\langle x(t)x(t+\tau)\rangle$. Now the problem of finding the velocity autocorrelation function reduces to calculating the position autocorrelation
function. The latter can be solved using diffusion theory \cite{MCG89,PET10}. For a spherical cell with radius $R$, it can be shown that:
\begin{eqnarray}
S_{x}(\omega_{0})
=4R^{2}\sum_{n}\frac{1}{x_{1n}^{2}(x_{1n}^{2}-2)}\frac{\frac{{x_{1n}^{2}}D}{R^{2}}}{(\frac{x_{1n}^{2}D}{R^{2}})^{2}+\omega_{0}^{2}},
\end{eqnarray}
where $x_{1n}$ are the zeros of the derivatives of spherical Bessel functions \cite{CAT88} and $D$ is the diffusion constant of the $^{3}$He gas. The relaxation time $T_{2}$ can be finally expressed as
\begin{equation}\label{eqnG2}
\Gamma_{2}=\frac{1}{T_{2}}=\frac{\alpha^{2}}{\hbar^{2}}\omega_{0}^{2}[S_{x}(\omega_{0})+S_{y}(\omega_{0})].
\end{equation}
Now using the data given in Ref.\cite{GEM10}, which used a radius of the spherical cell $R=3\textrm{cm}$ with $D=470\textrm{cm}^{2}\cdot\textrm{s}^{-1}$ and $\Gamma_{2}=8.0\times10^{-7}\textrm{s}^{-1}$ ($1\sigma$ value \cite{PET10}) into Eqn.(\ref{eqnG2}), we obtain an upper limit on $\alpha$ as
\begin{equation}\label{eqnA}
\alpha\leq6.9\times10^{-37}\textrm{kg}\cdot\textrm{m}\cdot\textrm{s}^{-1}.
\end{equation}
This upper limit is slightly more stringent than that given by Eqn.(\ref{eqn.aT1}) and so we are going to use $T_{2}$ to constrain the possible spin-velocity dependent new physics. However, the $T_{1}$ estimate  is simple and direct, and in some more complicated cell geometries one can get a constraint more easily and quickly by using the $T_{1}$ estimate.

\section{Constraining new interactions using the earth as a  source}
\subsection{ Vector-axial-vector interaction}
Taking advantage of the fact that there are about $10^{42}$ polarized electrons in the earth, Ref.\cite{HUN13} used the polarized electron spins of the earth as a source to constrain several possible exotic long-range spin-spin interactions. Inspired by this spirit, we notice that there are
$10^{51}$ nucleons in the earth. It might be even more advantageous to use the earth as a source to constrain some spin-velocity dependent interactions.
For example, for the vector-axial-vector interaction $V_{VA}(r)$ from the Lagrangian $\mathcal{L}_{X}=\bar{\psi}(g_{V}\gamma^{\mu}+g_{A}\gamma^{\mu}\gamma_{5})\psi X_{\mu}$,
this parity violating potential has the form:
 \begin{eqnarray}\label{eqnVA}
  V_{VA}(r)=\frac{\hbar g_{V}g_{A}}{2\pi}\frac{e^{-{r}/{\lambda}}}{r}\vec{\sigma}\cdot\vec{v},
  \end{eqnarray}
where $\lambda=\hbar/m_{X}c$ is the interaction range, $m_{X}$ the mass of the new vector boson and $g_{V}$,$g_{A}$ the vector and axial vector couplings.
$V_{VA}(r)$ is the Yukawa potential
multiplied by the $\vec{\sigma}\cdot\vec{v}$ factor.
Using the earth as a source, the Vector-Axial(V-A) potential generated by the earth on its surface is\cite{SUPL}:
\begin{eqnarray}\label{eqn.VA}
V_{VA}=\hbar g_{V}g_{A}\rho_{N}\lambda^{2}[(1-\frac{\lambda}{R_{\oplus}})e^{-\frac{\Delta y}{\lambda}}\nonumber\\
+(1+\frac{\lambda}{R_{\oplus}})e^{-\frac{2R_{\oplus}}{\lambda}}]\vec{\sigma}\cdot\vec{v},
\end{eqnarray}
where $\rho_{N}$ is the nucleon number density of the earth, $R_{\oplus}$ the earth radius, $\Delta y$ the probe to ground distance. If the force ranges are close to or smaller than the typical $\sim1\textrm{m}$ distance between the cell and ground,  the earth source can be considered as semi-infinite and Eqn.(\ref{eqn.VA}) is still valid as we will see later. 
If the new interaction $V_{VA}(r)$ exists, it will induce the $\alpha\vec{\sigma}\cdot\vec{v}$ interaction
on the earth surface  and affect the relaxation times of the spin polarized $^{3}$He gas. Plugging in $\rho_{N}=3.3\times10^{30}\textrm{m}^{-3}$ and using Eqn.(\ref{eqnG2}), we get
\begin{equation}\label{cgVgA}
g_{V}g_{A}\lambda^{2}[(1-\frac{\lambda}{R_{\oplus}})e^{-\frac{\Delta y}{\lambda}}+(1+\frac{\lambda}{R_{\oplus}})e^{-\frac{2R_{\oplus}}{\lambda}}]\leq 2.0\times10^{-34}\textrm{m}^{2}.
\end{equation}
\begin{figure}[htbp]\label{Fig.gVgA}
\centering
\includegraphics[scale=1.3, angle=0]{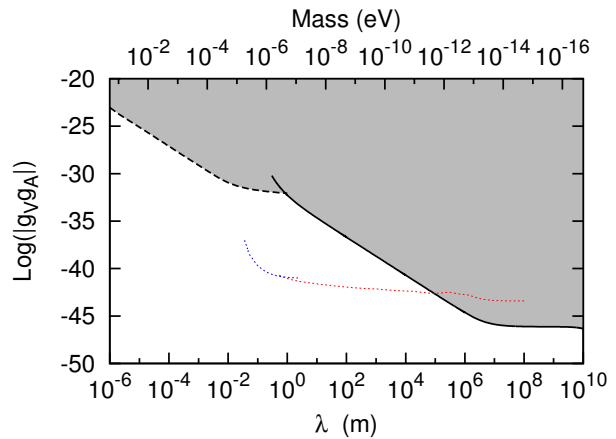}
 \caption{\small{(Color online)Constraint to the coupling constant product $|g_{V}g_{A}|$ as a function of the interaction range $\lambda$(new vector boson mass). The bold solid line
 is the result of this work; The dashed line is the result of  Ref.\cite{YAN13}; The blue and red dotted lines are the result of Ref.\cite{ADE13} which were
 derived by combining $g_{V}$ of Refs.\cite{SMI09,WAG12} with $g_{A}$ of Ref.\cite{VAS09} from separate experiments.The dark grey area is excluded by experiments
 of previous work \cite{YAN13} and this work, both directly constrain $|g_{V}g_{A}|$ in a single experiment. }}
\end{figure}
 The derived constraint on $g_{V}g_{A}$ is shown in FIG.1. For $\lambda\leq\Delta y$ much smaller than $R_{\oplus}$, the nonzero $\Delta y$ limits the practical force range. In this case Eqn.(\ref{eqn.VA}) can be approximated as:
\begin{equation}\label{eqn.VAG}
V_{VA}=\hbar g_{V}g_{A}\rho_{N}\lambda^{2}e^{-\frac{\Delta y}{\lambda}}\vec{\sigma}\cdot\vec{v},
 \end{equation}
 which is the same as the V-A potential derived in Ref.\cite{YAN14} for a plane plate in which the thickness goes to infinity.  
 
 In comparison with the previous result given by Ref.\cite{YAN13}, the constraint derived in this work has no improvement for ranges below $\sim1$m.
In the long distance limit with  $\lambda\gg R_{\oplus}$, Eqn.(\ref{cgVgA}) becomes:
\begin{eqnarray}
g_{V}g_{A}\leq7.4\times10^{-47}.
\end{eqnarray}
If the constraint derived from the neutron spin rotation experiment is extended to the long range $\sim10^{8}$m,  our work improves the existing direct experimental upper bound by as much as $\sim16$ orders of magnitude. When comparing with the results of Ref.\cite{ADE13}, which were derived by combining $g_{V}$ in Refs.\cite{SMI09,WAG12} and $g_{A}$ in Ref.\cite{VAS09}, we get $\sim3$ orders of magnitude improvement. The present work gives the best known constraint on $|g_{V}g_{A}|$ for ranges between $10^{-6}$m to $10^{8}$m. We emphasize that the limits on the vector-axial-vector interaction presented here are derived directly from a single laboratory experiment.

\subsection{Torsion induced by the earth on its surface}
Using neutron spin rotation in liquid helium, Refs.\cite{LEH14a,LEH14b} constrained possible \lq\lq  in-matter \rq\rq torsion for the first time. For torsion mentioned here, we refer to exactly the same type as in Refs.\cite{LEH14a,LEH14b}. These works showed that, for nonrelativistic motion of a spin in a torsion field, some time components of the torsion field couple to the spin through the form \cite{LEH14a} :
\begin{equation}
 V={\zeta}\vec{\sigma}\cdot\vec{v},
\end{equation}
which has exactly the same form as Eqn.(\ref{eqn.sdv}). $\zeta$ includes all other factors such as the distance, the source mass, {\it etc.}  As in Refs.\cite{LEH14a,LEH14b,KOS08}, here we only consider the leading order of the torsion background which is a constant. The torsion effect is considered to be very small and it is extremely difficult to be detected. Constraints on Earth-sourced torsion had been discussed in Ref.\cite{KOS08}, where rotating the apparatus or comparing the behavior of particles and antiparticles were proposed to detect torsion.These authors showed that the extensive body of sensitive experiments which have been conducted to search for CPT/Lorentz violation could be reinterpreted to constrain many components of a possible torsion tensor field in vacuum. We have found that relaxation time of the polarized $^{3}$He gas can constrain $\zeta$ without moving the apparatus.
Using the earth as a source and applying Eqn.({\ref{eqnA}}), it is easy to show that in natural units
\begin{equation}
\zeta=\alpha\leq1.3\times10^{-18}\textrm{GeV}.
\end{equation}
It might be more convenient to rewrite $\zeta$ as $\zeta=m_{S}c\xi$, where $m_{S}$ is the source mass. Now $\xi$ is dimensionless and its dependence on the source mass is isolated. Plugging in the earth mass $m_{\oplus}=5.97\times10^{24}\textrm{kg}$, we get
\begin{equation}
\xi=\frac{\alpha}{m_{S}c}\leq1.6\times10^{-70}.
\end{equation}

\section{other applications and proposed experiments} 
{\it Applying the method to the axial-axial-vector interaction} ---
It is possible to apply this analysis to other possible spin-velocity interactions. We have tried to constrain the axial-axial-vector interaction($V_{AA}$)  which also comes from $\mathcal{L}_{X}$. We find no improvement for ranges above $\sim 1$m in comparison with Ref.\cite{VAS09}. 

{\it Using the cell wall as a source}-- For $V_{VA}$ or $V_{AA}$ and ranges around $\sim1\textrm{mm}$, one might think to use the cell wall as a source, as done in Refs.\cite{POK10,PET10}. In this case $\alpha$ is not a constant anymore and cannot be moved out of the ensemble average. We would therefore need to calculate more complicated correlation functions such as $\langle\alpha[x(t),y(t),z(t)]\alpha[x(t+\tau),y(t+\tau),z(t+\tau)]v_{x}(t)v_{x}(t+\tau)\rangle$. The situation is simpler for $T_{1}$. The correlation time of velocities is 
$\sim10^{-10}\textrm{s}$, and on this time scale the atom can only move a distance $\sqrt{\langle v^{2}\rangle}\tau\sim10^{-7}\textrm{m}$. In this short distance for the force ranges under consideration, $\alpha(t)$ and $\alpha(t+\tau)$ can be treated as constants and moved out of the ensemble average. We have verified this approximation by Monte Carlo simulations. Using $T_{1}$, we obtain constraints for $|g_{V}g_{A}|$ $\sim3$ orders less stringent than Ref.\cite{YAN13} and for $g_{A}g_{A}$ more than $\sim1$ order less stringent than Ref.\cite{PIE12}. For $T_{2}$, the situation is much more complicated. If we approximate the spherical cell as a cube, and consider each wall as an infinite plate as in Ref.\cite{PET10}, we could get a constraint for $|g_{A}g_{A}|$. The result is also more than one order less sensitive than the existing constraint. This is not surprising since the relaxation times depend on the interactions through second order processes. At short distances, with a small amount of the source mass as the cell wall, it is hard for the relaxation time method to compete with the first order methods as in Refs.\cite{PIE12,YAN13,ZHE12,BUL13,CHU13}.

{\it Possibilities of applying the method to spin-spin-velocity dependent interactions}---
It is also possible to apply this method to search for many other types of spin-spin-velocity interactions. There are 6 different types of these interactions out of the 16 total derived in Ref.\cite{DOB06}. For these interactions, not only a relative velocity between the source and probe is required, but also both the probe and source have to be spin polarized. Thus they are even more difficult to search for experimentally. To the best of our knowledge, most of these interactions have never been experimentally searched for. The planned experiments and  projected constraints on such interactions were recently presented in Ref.\cite{LES14}. However, as in Ref.\cite{HUN13},  if we use polarized electrons in the earth as a source and the relaxation time method presented in this work, it would be possible to constrain these velocity dependent electron-spin-neutron-spin interactions for the first time. Though the relaxation time method is based on second order processes, it has advantages of using the earth as a large source and no need to move the apparatus.  The method presented in this work therefore opens a new path for probing various new interactions which are spin-spin-velocity dependent. The more complicated calculations needed to derive these constraints would require a more detailed consideration of the collisional spin dynamics in ensembles of polarized gas~\cite{WAL10} and will be the subject of future work. 
 
 {\it Possible improvement in sensitivity from $T_{1}$ }---
 We can try to further improve the sensitivity by dedicated experiments. Either for $T_{1}$ or $T_{2}$, the key is to constrain $\alpha$ better.  For $T_{1}$, in the limit that $1/\tau_{D}\ll\omega_{0}\ll1/\tau_{c}$ we find that $\alpha\propto\sqrt{{n}/{T_{1}}}$ after optimizing parameters. Since the measured $T_{1}$ is already as long as several thousand hours for polarized $^{3}$He with cell pressure of
$\sim1$ bar, it seems that the most promising way to improve the sensitivity is to reduce the gas number density. If $n$ is reduced to $\sim 1$ mbar, and the conditions for $T_{1}$ and $\omega_{0}$ remain the same as before, the sensitivity could be improved by a factor of $\sim30$ if the same long $T_{1}$ is observed. Furthermore, we notice that in Ref.\cite{PET10}, by subtracting the relaxation contributed from the magnetic field gradient, the residual $T_{2}$ from other sources was improved by about one order of magnitude from $\sim60\textrm{h}$ to $\sim500\textrm{h}$. Although a twice better $T_{2}$ was reported in Ref.\cite{ALL14},  no improvement in sensitivity can be made without knowing the magnetic field gradients. For $T_{1}$, according to our best knowledge, the contribution from the magnetic field gradient has not been subtracted yet. For $T_{1}$ measured as long as a few thousand hours, the relative magnetic field gradient, for example ${\partial_{x}{B}}/{B_{0}}$ is usually in order of $\sim10^{-4}\textrm{cm}^{-1}$. At this level, the gradient contributed relaxation is also in order of $\sim1000\textrm{h}$. There could be improvement for the $T_{1}$ method by precisely mapping the holding magnetic field. By reducing the gas pressure and mapping the field precisely, we might get an improvement in sensitivity by a factor of $\sim100$.

 {\it Possible improvement in sensitivity from $T_{2}$}---A similar result is obtained using $T_{2}$: $\alpha\propto\sqrt{{n}/{T_{2}}}$ after optimizing parameters. Since the cell pressure is already as low as a few mbars \cite{GEM10}, it is hard to imagine that sensitivity can be increased substantially by further reducing the pressure. Also the diffusion theory will break down for a much lower cell pressure, though the Redfield theory is still valid \cite{GOU10,SWA12}. On the other hand, lower cell pressure means fewer probing particles, a factor which will eventually dominate. The optimum gas pressure in this case is a theoretically interesting question. One might find the answer by using the theory presented in Refs.\cite{GOU10,SWA12}. For a $\sim$mbar cell,
if a thousand-hour-long $T_{2}$ could be observed the sensitivity could  be improved by one order of magnitude. In this case, since $\sim10^{-4}\textrm{cm}^{-1}$ field gradient is already realized for a holding field  as low as $\sim100\textrm{nT}$, it seems difficult though not impossible to improve the transverse relaxation time substantially. 
\section{conclusion}
By using the spin relaxation times of polarized $^{3}$He gas measured in previous experiments and  the earth as a source, we have constrained two types  of possible new interactions which are neutron spin-velocity dependent. We found that the best available $T_{2}$ relaxation times give slightly better constraints.
We derived new experimental limits on possible Vector-Axial type interactions with ranges from $\sim1$m
to $\sim10^{8}$m. At the distance of $\sim10^{8}$m, the limit is improved by $\sim16$ orders in magnitude in comparison with the previous  result of the neutron spin rotation experiment. In combination
with the previous result \cite{YAN13} which is more sensitive at short distances, we present the most stringent constraint derived directly from experiments
 on $g_{V}g_{A}$ ranging from $\sim10^{-6}$m to $\sim10^{8}$m (FIG.1). The methods presented in this work open up new possibilities to search for or constrain many possible spin-spin-velocity dependent interactions. By dedicated experiments, an improvement in sensitivity by a factor of $\sim100$ might be achieved using these ideas.
\begin{acknowledgements}
 We acknowledge support from the National Natural Science Foundation of China, under grant 11105128, 91126001, 11204032 and 51231002. We thank Dr. Alan Kosteleck\' y, W. M. Snow, H.Gao,  P.-H. Chu and Y.N. Pokotilovski for valuable discussions. We thank the anonymous referees for providing useful additional references. C.F. and H.Y. acknowledge partial support from the Indiana University Center for Spacetime Symmetries.
H.G. acknowledges support from grant No. SBK201241926.
\end{acknowledgements}

\end{document}